\documentclass[%
 aip,
 amsmath,amssymb,
 floatfix,
reprint,%
]{revtex4-2}

\usepackage{graphicx}
\usepackage{floatrow}
\usepackage{bm}
\usepackage{amsmath, amssymb}
\usepackage{color}
\usepackage{float}
\usepackage[sort&compress]{natbib}
\usepackage[normalem]{ulem}
\usepackage{color}
\usepackage{orcidlink}

\usepackage{algorithm}
\usepackage{algpseudocode}

\usepackage{times}

\usepackage{stackrel}

\usepackage{hyperref}
\hypersetup{
colorlinks=true,
citecolor=black,
linkcolor=black,
urlcolor=black
}

\newcommand{\R}{\mathbb{R}}

\newcommand{\p}{\mathbb{P}}

\newcommand{\fpt}{t_{\rm fh}}
\newcommand{\lap}[1]{\tilde{#1}}

\begin{document}

\title{Hitting the blinking target under stochastic resetting}

\author{Bartosz \.Zbik \orcidlink{0009-0002-6286-4499}}
\email{bartosz.zbik@student.uj.edu.pl}
\affiliation{Jagiellonian University, Krak\'ow, Poland}

\author{Bart{\l}omiej Dybiec \orcidlink{0000-0002-6540-3906} }
\email{bartlomiej.dybiec@uj.edu.pl}
\affiliation{Institute of Theoretical Physics
and Mark Kac Center for Complex Systems Research,
Faculty of Physics, Astronomy and Applied Computer Science,
Jagiellonian University, \L{}ojasiewicza 11, 30-348 Krak\'ow, Poland}

\author{Karol Capa{\l}a \orcidlink{0000-0002-8864-0760}}
\email{capala@agh.edu.pl}
\affiliation{Faculty of Computer Science, AGH University of Krakow, Mickiewicza~30, 30-059 Krak\'ow, Poland}
\affiliation{Academic Computer Center Cyfronet AGH, Nawojki~11, Krak\'ow, 30-950, Poland}

\author{Zbigniew Palmowski \orcidlink{0000-0001-9257-1115}}
\email{zbigniew.palmowski@pwr.edu.pl}
\affiliation{Department of Applied Mathematics,
Wroc{\l}aw University of Science and Technology, Wybrze\.ze Wyspia\'nskiego~27
50-370 Wroc{\l}aw, Poland}

\author{Igor M. Sokolov \orcidlink{0000-0002-4688-9162}}
\email{igor.sokolov@physik.hu-berlin.de}
\affiliation{Institute of Physics, Humboldt University of Berlin, Newtonstrasse 15, D-12489 Berlin}

\date{2026-02-20}

\begin{abstract}
The first hitting times of a stochastic process, i.e., the first time a process reaches a particular level, are of significant interest across various scientific disciplines, including biology, chemistry, and economics. We modify the standard setup by allowing the target to spontaneously switch between two states, either active or inactive, and investigate the distribution of first hitting times accrued while the target is active.
For this setup, we provide closed formulas for the distribution of the first hitting time.
Additionally, we can introduce stochastic resetting to the underlying process and, utilizing our results, derive the formulas for the first time the active target is hit by the process under stochastic resetting.
Interestingly, we show that resetting in this setup still leaves some memory; the system is no longer Markovian, which prevents a straightforward application of standard techniques. 
The analytical results are accompanied by computer simulations of Langevin dynamics.
\end{abstract}


\maketitle

\begin{quotation}
Stochastic resetting and first-passage processes are fundamental concepts in the theory of stochastic dynamics, with numerous applications in physics, chemistry, and biology. These two notions are closely related, as stochastic resetting can significantly reduce the mean first-passage time by terminating inefficient trajectories. In this work, we investigate the first-passage problem under stochastic resetting in the presence of gated target kinetics. The target undergoes dichotomous switching between two configurations, with detection possible only in one of them, introducing an additional layer of complexity to the search dynamics.
We  derive closed-form analytical expressions for the distribution of first-passage times and the mean first-passage time, which are compared with stochastic simulations.
These findings have potential implications for a broad class of search processes, including molecular recognition, enzymatic reactions, and transport phenomena in heterogeneous environments.

\end{quotation}

\section{Introduction \label{sec:introduction}}
First passage time~\cite{redner2001} is a key quantity characterizing various processes. 
It can be associated with deterministic actions---such as identifying who crosses the finish line first---or with stochastic processes involving randomness, such as regulatory proteins or molecules reaching a critical threshold level. 
Typically, first passage time events are studied in the context of stochastic processes \cite{vankampen1981}. The mean first-passage time (MFPT), which is the average of all first passage times, introduces a characteristic timescale for a stochastic event to occur for the first time.

First passage time can be studied in the context of Brownian motion (BM). 
For example, one can calculate the average time required for a diffusing particle to exit a finite domain. 
In cases of normal diffusion and superdiffusion, this process is characterized by a finite MFPT~\cite{getoor1961,widom1961stable,kesten1961random,kesten1961theorem,zoia2007}.
However, the situation differs for subdiffusion, where trapping events may cause a particle to remain indefinitely within the domain. 
Analogously, expanding the domain of motion can render the MFPT ill-defined. 
For instance, in one dimension, if one boundary of the interval is moved toward infinity, the MFPT diverges and can no longer characterize escape kinetics \cite{dybiec2016jpa}. 
This divergence arises because the meandering trajectory has sufficient space to stray far from the boundary. 
Although the escape process from a half-line cannot be described by MFPT, the first passage time density remains well-defined. 
For Brownian motion, it follows the L\'evy-Smirnov distribution, which is a one-sided, heavy-tailed distribution \cite{metzler2004,chechkin2006}. 
The finite MFPT can be reintroduced by incorporating stochastic resetting \cite{evans2011diffusion-jpa,gupta2022stochastic}, which restarts inefficient meandering trajectories. 
As a result, the first escape from a half-line under Poissonian resetting is characterized by a finite MFPT.

In this manuscript, we study a problem that, in some sense, resembles an escape from a half-line. 
We assume a continuous process on the real line that must hit a specific position, which additionally must be in an active state. 
This makes the target even harder to reach than in the standard half-line escape problem. 
If the target is inactive, the particle may pass to the half-line located on the opposite side of the target. 
Consequently, the dynamics governing the target’s evolution introduce an additional slowdown to the escape process. 
Nevertheless, stochastic resetting can be employed to facilitate the search kinetics. Resetting reduces the likelihood that the process strays far from the target. 
Thanks to this mechanism, even in the presence of target dynamics, the average search time to hit a blinking target can remain finite.

Models in which the target is accessible only for a finite duration are of fundamental importance across numerous scientific disciplines.
In the theory of chemical reactions, one considers problems where a particle must not only be located in the appropriate position but also be in the proper internal state for a chemical reaction to occur.
In this context, such systems are referred to as gated reactions or, more generally, gated targets~\cite{McCammon1981,Szabo1982}.
Alongside continuous models, discrete representations of chemical reactions have also been studied~\cite{rojo2011intermittent,Scher2021a,Scher2021b}.
A popular example is the binding of ligands to proteins, where the proteins are treated as targets that can exist in either an active or inactive state. For a ligand to bind, it must not only reach the protein but also encounter it in its active form~\cite{McCammon1981,Northrup1982,Spouge1996}.
This problem has been extensively studied for various combinations of activity and mobility of both ligands and proteins~\cite{Zhou1996,Berezhkovskii1997,Makhnovskii1998,Benichou2000,Gopich2016}.
The model has also been extended to describe search processes in two and three dimensions~\cite{Szabo1980,Bressloff2015a}.
The same model can also be applied to investigate the transient kinetics of fluorescence quenching in through-solvent photoinduced electron transport reactions~\cite{Bandyopadhyay2000}.

The formalism of chemical reactions \cite{Weiss1986} is also used to describe the translocation of proteins through cellular pores.
Molecules can either pass through the pore -- representing a successful encounter with the target -- or become trapped.
The forces driving these outcomes typically vary over time, allowing the molecule to relax between active translocation phases.
This relaxation corresponds to a resetting process, whereas actual translocation occurs only during periods when a pulling force acts across the pore~\cite{narsimhan2016translocation, szymczak2016periodic, capala2022stochastic}.
Such intermittent activation effectively generates a gated target.
Gated kinetics\cite{grebenkov2024target} have been studied for BM without \cite{mercado2019firsthitting,scher2024continuous,szabo1984localized} and with resetting\cite{mercado2021search} also in the presence of drift\cite{biswas2023rate}.
These studies can rely on application of the Smoluchowski-Fokker-Planck equation \cite{mercado2019firsthitting,mercado2021search} or general properties of gated Markov processes \cite{szabo1984localized,scher2024continuous}.
Our approach generalizes the one of Refs.~\onlinecite{szabo1984localized,scher2024continuous} to the case with stochastic resetting which provides a general framework which also allows to reproduce results of Refs.~\onlinecite{mercado2021search,biswas2023rate} as special cases.

The concept of gated targets also naturally arises in the study of search processes.
A classical example is animal foraging, where the sought-after resource may be available only intermittently -- for instance, when prey does not hide in its refuge~\cite{hahn2005skuas, suraci2013short}.
Analogous considerations appear in problems of electronic signal detection, where one seeks to determine the first time a static receiver captures the signal of a transmitter within its detection range~\cite{inaltekin2007event, song2011time}.
In this context, the gating mechanism can model the intermittent operation of electronic devices due to power constraints or hardware failures~\cite{kumar2023inference}.

The escape problem involving a gated target has been the subject of extensive theoretical investigation \cite{Godec2017,Mercado2021firsthitting, Toste2022}, particularly in the case of escape from a potential well, where it has been demonstrated that the influence of a blinking barrier depends on the local gradient of the potential in its vicinity~\cite{Bressloff2015b,Bressloff2015c}.
A particularly active research area involving analogous problems is that of quantum walks with resetting, where the impossibility of continuous measurement naturally gives rise to a gated target~\cite{mukherjee2018quantum, rose2018spectral, yin2023restart, yin2025restart}.

The current studies are the closest in scope to the ones presented in  Ref.~\onlinecite{Bressloff2020}, discussing stochastic resetting combined with a gated target for the case of BM. In both models, the one of Ref.~\onlinecite{Bressloff2020}, and our one, the target switches in a Markovian manner between active and inactive states at given rates, and can only be hit when active. Our model, however,  differs significantly from the one explored in Ref.~\onlinecite{Bressloff2020}. The first difference concerns the nature of the inactive state, which is reflective in Ref.~\onlinecite{Bressloff2020} and transparent in our model, so that in one dimension the searcher can pass the target, and hit it later from the opposite side.
For the unbiased Brownian motion on a line, due to the reflection property, the distinction between transparent and reflecting boundaries does not affect the statistics of first‑hitting times.
For other stochastic processes, e.g., the Ornstein–Uhlenbeck process, it leads to fundamentally different models.
The question, from which side of the target is hit, stays however meaningful also for the unbiased Brownian motion, and is investigated in simulations. 
Moreover, in Ref~\onlinecite{Bressloff2020} the state of the target after the resetting event is renewed, and chosen randomly from the equilibrium distribution, while in our setup the the dynamics of the target is unaffected by the stochastic resetting. For this model, we provide analytical results for the first passage time distribution, and for the mean first passage time.
Analytical results are  compared with results of numerical simulations.

The work is organized as follows:
The Model section (Sec.~\ref{sec:model}) introduces the model and its components in a comprehensive manner (Sec.~\ref{sec:modelformulation}) and in a more detailed and rigorous way (Sec.~\ref{sec:modelformalization}).
The Results section (Sec.~\ref{sec:results}) presents the main findings and is divided into two main parts: Analytics (Sec.~\ref{sec:analytics}) and Numerics (Sec~\ref{sec:langevin}).
The first part of the Results section presents analytical results and derived formulas,  while the second part covers the numerical approach based on the Langevin dynamics.
The Results section ends with a comparison of the results obtained from both parallel approaches (Sec.~\ref{sec:comparison}).
The paper finishes with the Summary and Conclusions (Sec.~\ref{sec:summary}).

\section{Model\label{sec:model}}

We start with a comprehensible description of the model (Sec.~\ref{sec:modelformulation}), which is later formalized (Sec.~\ref{sec:modelformalization}).
This structure allows us to present the model together with the main analytical findings more effectively in Sec.~\ref{sec:results} (Results).

\subsection{Model formulation\label{sec:modelformulation}}

\begin{figure}[h]
    \centering
    \includegraphics[width=0.7\linewidth]{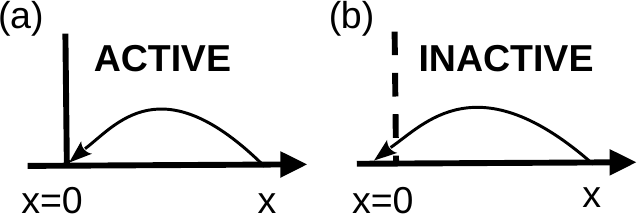}\\
    \includegraphics[width=0.9\linewidth]{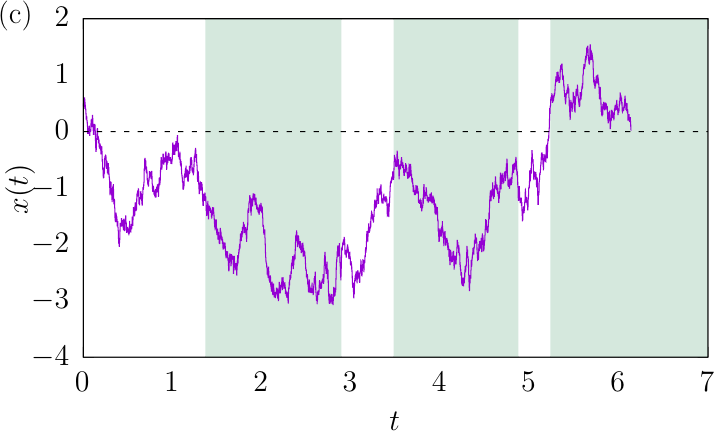}
    \caption{Experimental setup corresponding to hitting an active target (top left panel --- (a)) and passing over inactive target (top right panel --- (b)). 
    The bottom panel (c) presents a sample trajectory, green area show when target is in active state.
    }
    \label{fig:setup}
\end{figure}

We utilize and further extend a classical paradigmatic model of Brownian motion \cite{gardiner1983}.
A particle performing a Brownian motion on the real line is described by the following Langevin equation
\begin{equation}
    \dot{x}(t)=\xi(t),
    \label{eq:langevin}
\end{equation}
with $x(0)=x_0 > 0$.
In Eq.~(\ref{eq:langevin}) $\xi(t)$ represents the Gaussian white noise
\begin{equation}
\langle \xi(t) \rangle= 0 \;\;\;\;\; \mbox{and} \;\;\;\;\; \langle \xi(t)  \xi(s) \rangle= \sigma^2 \delta(t-s).    
\end{equation}
The BM continues until $x(t)$ crosses the origin, i.e., the random walker hits the target placed at $z \equiv 0$.
Additionally, the target switches between two states: active ($A$) and inactive ($I$).
The process is terminated only if the state of the hit target is active, see Fig.~\ref{fig:setup}(a).
If the target is inactive, a process $x(t)$ can move to the opposite side of the origin, see Fig.~\ref{fig:setup}(b).
Such a process resembles the process of escape from a half-line; nevertheless, due to target dynamics, not every passage over the origin is considered a hit, see Fig.~\ref{fig:setup}(c), which further slows the escape dynamics.
In Fig.~\ref{fig:setup}(c), the green area indicates when the target is in an active state. 
Therefore, the first and second crossings over the origin are not considered hits, since the target was in the inactive state.
In contrast to the escape from the half-line, the target can be hit both from the right and from the left, because a particle can cross the origin if the target is in the inactive state.
Consequently, the particle dynamics is defined on the whole real line.

In order to reintroduce a finite MFPT, the Brownian motion is affected by Poissonian resetting with a reset rate $r$.
The distribution of time intervals between two consecutive resets follows the exponential density 
\begin{equation}
\phi(t) = r \exp(-rt),    
\end{equation}
where $r$ is the (fixed) reset rate.
The mean time between two consecutive restarts is $\langle t \rangle = 1 / r$.
Stochastic resetting restarts the motion from $x_0$; however, it does not affect the target dynamics, while in~\cite{Bressloff2020} it is assumed that resetting also affects the target dynamics, i.e., at resetting events the target state is generated from the equilibrium distribution.

The random motion is affected by the target, which is placed at $x=0$.
Its state is described by the Markovian dichotomous process $J(t)$, see Ref.~\onlinecite{horsthemke1984}.
It takes two states: $A$ and $I$, which are considered active and inactive states
\begin{equation}
    \label{eq:dn}
    A \stackrel[\beta]{\alpha}{\rightleftarrows} I,
\end{equation}
while $\alpha$ and $\beta$ are switching rates.
The target can only be found if it is in the active state; thus, the inactive state can be considered as the state in which the target is invisible or hidden.
The initial state of the target is either deterministic or drawn from a known two point probability distribution $\p(J(0)=A)=1-\p(J(0)=I)$ with fixed $\p(J(0)=A)$, e.g., $\p(J(0)=A) = \pi_A$.

The central quantity of interest is the first passage time $\tau$, which is the time of first hitting the target (origin) when the target is in the active state.
Properties of first passage times are determined by the first passage time density, which in turn is influenced by the initial condition, target position, target dynamics, and stochastic resetting (if applied).
From first passage times, it is possible to calculate the mean first passage time as the ensemble average.
However, not every escape process can be characterized by a finite MFPT.

The studied model differs from the escape from the  half-line.
Nevertheless, in some special limits, the model can be equivalent to the escape from a half-line.
Therefore, we recall basic information about the properties of escape kinetics from a half-line.
For a particle performing BM on the real line, the first passage time density from $x_0 \to z$ is given by the L\'evy-Smirnov distribution
\begin{equation}
    f(t|x_0,z)= f(t|x_0-z, 0)=   \frac{|x_0 - z|}{\sqrt{2\pi\sigma^2 t^3} } \exp \left[ - \frac{(x_0 - z)^2}{2\sigma^2 t} \right],
    \label{eq:fpt-hl}
\end{equation}
which has the $t^{-3/2}$ asymptotics.
The cumulative distribution function reads
\begin{equation}
    \mathcal{F}(t|x_0,z) = \mathcal{F}(t|x_0-z,0) = \mathrm{erfc} \left[ \frac{|x_0- z|}{\sqrt{2 t} \sigma }  \right],
    \label{eq:cdf-hl}
\end{equation}
where $\mathrm{erfc}(.)$ is the complementary error function
($\mathrm{erfc}(z)=1-\mathrm{erf}(z)=1-\frac{2}{\sqrt{\pi}}  \int_0^z e^{-s^2}  ds$).
The cumulative distribution function has $t^{-1/2}$ asymptotics.
The MFPT diverges, but stochastic resetting can render it a finite value, which reads~\cite{evans2011diffusion}
\begin{equation}
    \langle \fpt^R \rangle =  \frac{1}{r} \left[ \exp \sqrt{  \frac{2r (x_0-z)^2 }{\sigma^2}  }  -1  \right].
    \label{eq:mfpt-sr}
\end{equation}

\subsection{Model Formalization \label{sec:modelformalization}}

For the analytical derivation, it suffices to consider a pair of independent, continuous time Markov processes $X(t)$ and $J(t)$, where $X(t) \in \R$ represents the position of a particle at time $t$, and $J(t) \in \{ A, I \}$ represents the state of the target, either active ($A$) or inactive ($I$).

The process $X(t)$ is assumed to have continuous trajectories, for instance, it can be a Brownian motion (Wiener process), but other options are also possible, and is described by a transition density
\[ p(t, y|x) dy = \p(X(t) \in [y, y + dy] | X(0) = x_0) . \]
If not stated otherwise, we assume that the process starts at a fixed position $x_0$ (i.e., $X(0) \equiv x_0$).

The process $X$ is strictly related with the first passage time
\begin{equation}
\fpt := \inf \{ t \geqslant 0 : X(t) = z \},
\end{equation}
denoting the first time, when the process $X$ arrives at a fixed point $z$.
We will denote the density of $\fpt$ by $f(t|x_0)$.

The process $J(t)$ is a two-state Markov chain, as in Eq.~(\ref{eq:dn}), with transition probabilities from $i \to j$ denoted by $P_{i\to j}^J(t)$:
\begin{equation}
P_{i \to j}^J(t) := \p(J(t) = j | J(0) = i).
\end{equation}
We will also use the following notation: $\lambda := \alpha + \beta$,  $\pi_A:=\beta / \lambda$ and $\pi_I:=\alpha / \lambda$, where $\pi_A$  and $\pi_I$ are probabilities to find a target in active or inactive state in equilibrium see Ref.~\onlinecite{horsthemke1984}.

Our first aim is to study the distribution of a stopping time
\begin{equation}
\label{eq:tau-def}
\tau := \inf \{  t \geqslant 0 : X(t) = z ~~~\text{and}~~ J(t) = A \},
\end{equation}
which represents the first time the process $X$ arrives at the fixed target position $z$ ($X(t) = z$) while the target is active ($J(t) = A$).
We will denote the density of $\tau$ by $g(t|x_0, j_0)$ where $j_0$ is the initial state of the target ($J(0) = j_0$) and $X(0) \equiv x_0$.

We can further extend the model by adding stochastic resetting to the process $X(t)$; the process $J(t)$ is unaffected by resetting.
We will use $X^R(t)$ to denote the process under Poissonian resetting, $r$ for the reset rate, and the superscript $^R$ to denote quantities defined analogously to the process without resetting.
In particular, the analog of $\tau$ for $X^R(t)$ would be $\tau^R$ defined as
\begin{equation}
\label{eq:tauR-def}
\tau^R := \inf \{  t \geqslant 0 : X^R(t) = z ~~~\text{and}~~ J(t) = A \}.
\end{equation}
The density of $\tau^R$ we will denoted by $g^R(t|x_0, j_0)$.

\section{Results \label{sec:results}}

The problem of hitting the blinking target can be described within the framework of first passage time density (Sec.~\ref{sec:analytics}).
The main results of the analytical approach are the Laplace transforms of $\tau$ densities, see definition in Eq.~(\ref{eq:tau-def}), which in the absence of resetting are provided in Sec.~\ref{sec:results-no-resetting} by formulas~(\ref{eq:DqxA}) and~(\ref{eq:DqxI}).
Under stochastic resetting $\tau^R$ is defined by Eq.~(\ref{eq:tauR-def}), while main results are given in Sec.~\ref{sec:results-resetting} by Eqs.~(\ref{eq:DRqx0A}) and~(\ref{eq:DRqx0I}).

In parallel to the analytical description, the properties of the model can be studied by simulating the Langevin equation (Sec.~\ref{sec:langevin}).
Finally, the results obtained by both methods are compared (Sec.~\ref{sec:comparison}).

\subsection{Analytics\label{sec:analytics}}

The main problem solved in this work is how to derive the statistics of $\tau$, based on the knowledge of $X$, $\fpt$ and $J$ (or equivalently $p(t, x|x_0)$ -- propagator of $X(t)$, $f(t|x_0)$ -- density of $\fpt$ and $P^J_{i\to j}(t)$ -- transition probabilities of $J(t)$).
The results are presented in Sec.~\ref{sec:results-no-resetting}.
A similar question was asked in~\cite{kumar2023inference}, but there, the authors derived the statistics of $\fpt$ based on $\tau$.

Furthermore, in Sec.~\ref{sec:results-resetting}, we enrich the model with stochastic resetting and derive the statistics of $\tau^R$ based on the results from Sec.~\ref{sec:results-no-resetting}.
The procedure for going from $\fpt$ to $\fpt^R$ is well known~\cite{pal2017first, chechkin2018random}, but in our setup under resetting, going from $\tau$ to $\tau^R$, the processes $X^R(t)$ and $J(t)$ are no longer independent; thus, the standard methods require further adjustments.

\subsubsection{The Case Without Resetting \label{sec:results-no-resetting}}
Consider a particle starting at $x_0$ at time 0.
Let us specify two parts that contribute to the density of $\tau$.
The first comes from the case where $\tau = \fpt$, that is, the first time the particle arrived at $z$, the target was active $J(\fpt) = A$.
In the other case $J(\fpt) = I$, so we know the exact state of the system at time $\fpt$: $(X(\fpt) = z, J(\fpt) = I)$. Combining these two cases and using the Markov property, we can write the following equation~(see~\cite[Eq.~(1)]{kumar2023inference}) for the density of $\tau$
\begin{eqnarray}
\label{eq:Dtx0j0}
g(t|x_0, j_0) & = & f(t| x_0) P^J_{j_0 \to A}(t) + \\ \nonumber
&& + \int_0^t f(t'| x_0) P^J_{j_0 \to I}(t') g(t-t'|z, I) {\rm d}t'~.
\end{eqnarray}
In the above equation, the only unknown part on the right hand side is $g(t-t'|z, I)$.
Therefore, the problem would be solved if we find $g(t|z, I)$.
To do so, we now assume that the system starts at $X(0) = z,~J(0)=I$ and apply the Markov property at the first instance of the target switching to $A$
\begin{equation}
\label{eq:DtmI}
g(t|z, I) =  \int_0^t \beta e^{-\beta t'} \left[  \int_{-\infty}^{\infty} p(t',x | z) g(t - t'|x ,  A) {\rm d}x \right] {\rm d}t'.
\end{equation}
In the above equation, we can replace $g(t - t'|x,  A)$ using Eq.~(\ref{eq:Dtx0j0}), which gives a closed formula for $g(t|z, I)$.
The resulting triple integral equation can be solved by applying the Laplace transform ($\lap f(q) := \mathcal{L}\{ f \} (q) := \int_0^\infty e^{-q t} f(t) {\rm d}t$) with respect to the variable $t$.
A part of derivation  is presented in Appendix~\ref{sec:appendix-derivation}.
From these calculations, it turns out that it is useful to define the following expressions
\begin{eqnarray}
\label{eq:I1q}
\lap{I_1}(q) & = & \beta \int_{-\infty}^{\infty} \lap p(q + \beta, x | z) \times \\ \nonumber
 && \times \left  [\pi_A \lap f(q|x) + \pi_I \lap f(q + \lambda|x) \right ]{\rm d}x,
\end{eqnarray}
\begin{eqnarray}
\label{eq:I2q}
\lap I_2(q) & = &  \beta \int_{-\infty}^{\infty} \lap p(q + \beta, x | z) \times \\ \nonumber
&& \times  \pi_I \left  [ \lap f(q|x) - \lap f(q + \lambda|x) \right ]{\rm d}x.
\end{eqnarray}
This allows us to express $\lap g(q|z, I)$ in a compact form
\begin{equation}
\label{eq:DqmI}
\lap g(q|z, I) = \frac{ \lap{I_1}(q) }{1 - \lap{I_2}(q) }.
\end{equation}
Calculating the Laplace transform of Eq.~(\ref{eq:Dtx0j0}) and using Eq.~(\ref{eq:DqmI}), we obtain exact formulas for the Laplace transform of densities of $\tau$
\begin{equation}
\label{eq:DqxA}
\begin{aligned}
\lap g(q| x_0, A) = \pi_A \lap f(q| x_0) + \pi_I \lap f(q + \lambda|x_0) + \\
\pi_I \left [\lap f(q| x_0)  -  \lap f(q + \lambda|x_0) \right ]  \frac{ \lap{I_1}(q) }{1 -  \lap{I_2}(q) }
\end{aligned}
\end{equation}
and
\begin{equation}
\label{eq:DqxI}
\begin{aligned}
\lap g(q| x_0, I)  = \pi_A  ( \lap f(q| x_0) - \lap f(q + \lambda|x_0) ) + \\
\left [\pi_I \lap f(q| x_0) + \pi_A \lap f(q + \lambda|x_0) \right ] \frac{ \lap{I_1}(q) }{1 - \lap{I_2}(q) }.
\end{aligned}
\end{equation}
Following Ref.~\onlinecite{kumar2023inference}, one can obtain an even simpler formula, assuming that the initial state of the target is equilibrium $E$ (meaning $\p(J(0) = A) = \pi_A$ and $\p(J(0) = I) = \pi_I$)
\begin{equation}
\label{eq:DqxE}
\lap g(q| x_0, E)  = \lap f(q| x_0) \left \{ \pi_A + \pi_I  \frac{ \lap{I_1}(q) }{1 - \lap{I_2}(q) } \right \}.
\end{equation}
The asymptotic behavior of formulas (\ref{eq:DqmI}) -- (\ref{eq:DqxE}) depends on the specific characteristics of the underlying process. Consequently, it is not possible to draw general conclusions since each case requires a separate, tailored examination.

\subsubsection{The Case With Resetting \label{sec:results-resetting}}
Changing the setup to the case with Poissonian resetting, we are interested in the distribution of $\tau^R$, defined in Eq.~(\ref{eq:tauR-def}).
We start by looking at the Laplace transform of $\tau^R$ and considering two parts contributing to $\tau^R$: the first, when the particle reaches the target before the first resetting event, and the complement event
\begin{eqnarray}
\lap g^R(q|x_0, j_0) & = &  \langle e^{-q \tau^R} \rangle_{j_0} \\ \nonumber
& = &  \langle e^{-q \tau} | \tau \leqslant R \rangle_{j_0}  \p(\tau \leqslant R)+ \\ \nonumber
&& +  \langle e^{-q (R + \tau^R_1)} | \tau > R \rangle_{j_0}  \p(\tau > R),
\end{eqnarray}
where $\langle A| B\rangle_{j_0}$ denotes the expected value of $A$ under condition $B$,
the lower index $j_0$ denotes the initial state of the target, $R$ stands for the time of the first reset,
and $\tau^R_1$ is defined as
\[ \tau^R_1 = \inf \{ t > R : X^R(t) = z ~~~\text{and}~~ J(t) = A \} - R. \]
In the standard case of stochastic resetting~\cite{pal2017first, chechkin2018random}, $\tau^R_1$ turns out to be an independent copy of $\tau^R$; however, this is not true in our setup, as after a reset, the system still has some memory about the state of the target.
To overcome this issue, one can condition the expression $\langle e^{-q (R + \tau^R_1)} | \tau > R \rangle_{j_0}$ on the state of the target
\begin{equation}
\label{eq:DRqx0j0}
\begin{aligned}
\lap g^R(q|x_0, j_0) = \langle e^{-q \tau^R} \rangle_{j_0} = \langle e^{-q \tau} | \tau \leqslant R \rangle_{j_0}  \p(\tau \leqslant R)+ \\
 \sum_{i \in \{I, A\}} \langle e^{-q (R + \tau^R_1)} | \tau > R, J(R) = i  \rangle_{j_0}  \p(\tau > R, J(R) = i).
\end{aligned}
\end{equation}
Then assuming the knowledge of $\lap g(q|x_0, j_0)$ calculated in the previous section one can derive a closed formula for the Laplace transform of $\tau^R$
\begin{equation}
\label{eq:DRqx0A}
\begin{aligned}
\lap g^R(q|x_0, A) = \\
\frac{(r (\beta +q)+q (\lambda +q)) \lap g(r+q|x_0,A)+\alpha  r
   \lap g(r+q|x_0,I)}{r (\beta +q) \lap g(r+q|x_0,A)+\alpha  r
   \lap g(r+q|x_0,I)+q (\lambda +q)},
\end{aligned}
\end{equation}
\begin{equation}
\label{eq:DRqx0I}
\begin{aligned}
\lap g^R(q|x_0, I)  = \\
   \frac{\beta  r~\lap g(r+q|x_0,A)+((r+q) (\alpha +q)+\beta  q)~ \lap g(r+q|x_0,I)}{r
   (\beta +q)~\lap g(r+q|x_0,A)+\alpha  r~\lap g(r+q|x_0,I)+q (\lambda +q)}.
\end{aligned}
\end{equation}
Note that both Eqs.~(\ref{eq:DRqx0A})~and~(\ref{eq:DRqx0I}) depend on both $\lap g(q|x_0, I)$ and $\lap g(q|x_0, A)$.

It is worth noting that to derive these formulas it was necessary to solve a system of two coupled algebraic equations, namely Eq.~(\ref{eq:DRqx0j0}) for $j_0 = A$ and $j_0 = I$, which differs from the case of the standard setup for stochastic resetting with just one equation.
Even assuming the initial state of the target to be equilibrium, $E$, the formula still cannot be expressed only with $\lap g(q|x_0, E)$
\begin{equation}
\label{eq:DRqx0E}
\begin{aligned}
\lap g^R(q|x_0, E) = \\
\frac{ (r+s) (\lambda +s)~\lap g(r+q|x_0,E) }{
 \left(r s~\lap g(r+q|x_0,A) + r \lambda~\lap g(r+q|x_0,E) + s (\lambda +s) \right)}
\end{aligned}
\end{equation}

The access to Laplace transforms $\lap g^R(q|x_0,j_0)$ allows, through differentiation, to calculate the mean first hitting time $\langle \tau^R \rangle_{j_0}$:
\begin{align}
\label{eq:mean-tauR-A}
\langle \tau^R \rangle_A &= \frac{1}{r}\frac{\left(1 - \lap g(r|x_0,A) \right)}{ \lap g(r|x_0,E) } \\
\label{eq:mean-tauR-I}
\langle \tau^R \rangle_I &= \frac{r ~\lap g(r|x_0,A)-(\lambda +r) ~\lap g(r|x_0,I)+\lambda }{r \lambda ~\lap g(r|x_0,E)}.
\end{align}
Note that if $\langle \tau \rangle_{j_0} < \infty$ then the asymptotic behavior of $\langle \tau^R \rangle_A$ as $r \to 0$ (limit of no resetting) is $- \lap g'(q =0|x_0,A) \equiv \langle \tau \rangle_A$, which is exactly the expected asymptotic behavior.
Similarly, $\langle \tau^R \rangle_I \to - \lap g'(q=0|x_0,I)$ as $r \to 0$.

\subsubsection{Application to the case of Brownian Motion \label{sec:results-BM}}
\begin{figure}[h]
\centering
\includegraphics[width=\textwidth]{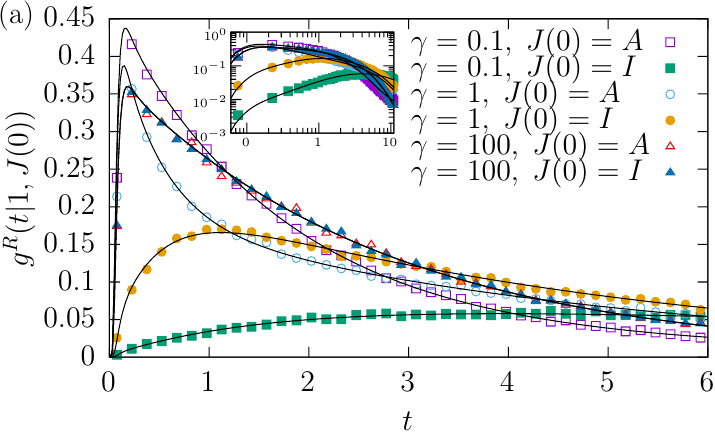}\\
\includegraphics[width=\textwidth]{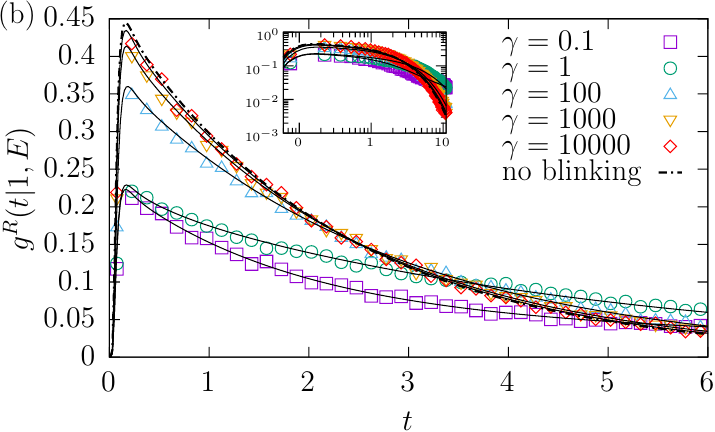}\\
\caption{Densities $g^R(t|x_,J(0))$ of $\tau^R$  with $J(0)=A$ and  $J(0)=I$ for various switching rates $\gamma$ (top panel -- (a)) and densities for the equilibrium initial condition $g^R(t|x_,E)$ (bottom panel -- (b)).  Solid lines corresponds to the analytical solution (see Sec.~\ref{sec:results-BM}) while points represent results obtained from simulation of Langevin equation (see Sec.~\ref{sec:langevin}). Model parameters: $z=0, x_0 = 1, \sigma = \sqrt{2}$ and $r=10$.
Insets replot the main plots in the log-log scale.}
\label{fig:density}
\end{figure}

To verify the theoretical results from Secs.~\ref{sec:results-no-resetting} and~\ref{sec:results-resetting}, we applied the theory to an explicitly known setup of Brownian motion (BM) in 1D.
Moreover, we assume that the switching rates controlling the state of the target are the same, i.e., $\alpha=\beta$ and denoted with $\gamma$, i.e., $\gamma=\alpha=\beta$.
For the chosen process, both the propagator $p(t, x|x_0)$ and the distribution $f(t|x_0)$ of $\fpt$ are well known; see, e.g.,~Ref~\onlinecite[Eq. 2.0.2]{borodin2002handbook} 
\begin{equation}
    p(t, x|x_0) = \frac{1}{\sqrt{2 \pi  \sigma^2 t}} e^{-\frac{(x - x_0)^2}{2 \sigma^2 t}},
\end{equation}
and
\begin{equation}
    f(t|x_0) = \frac{|x_0 - z|  e^{-\frac{(x_0 - z)^2}{2 \sigma ^2 t}}}{\sqrt{2 \pi } \sigma 
   t^{3/2}}.
\end{equation}
This allowed us to explicitly calculate the distribution of $\tau$ and $\tau^R$ in the Laplace domain.
The results without resetting ($\tau$) agree with formula obtained for this particular setup using other methods~Ref.~\onlinecite[Eq.~(10) therein]{mercado2019firsthitting}, see also Appendix~\ref{sec:bmapp}.
The exact density of $\tau^R$ is presented in Fig.~\ref{fig:density}. 
The top panel of Fig.~\ref{fig:density} compares the exact densities of $\tau^R$ depicted with solid lines with various switching rates $\gamma$ and different initial conditions of the target: $J(0)=A$ and $J(0)=I$.
For small $\gamma$, e.g., $\gamma=0.1$ or $\gamma=1$, densities with different initial  target states are distinct.
With the increasing switching rate $\gamma$, they become similar, and for sufficiently large $\gamma$, e.g., $\gamma=100$, they are indistinguishable because the states of the target change rapidly.
In the bottom panel of Fig.~\ref{fig:density}, exact densities with the equilibrium initial condition ($\p(J(0) = A) = \p(J(0) = I) = \frac{1}{2}$) are presented.
For such an initial condition $g(t|x_0, E) = \frac{1}{2} \left[ g(t|x_0, A) + g(t|x_0, I) \right] $.
Moreover, as it is visible from the bottom panel of Fig.~\ref{fig:density}, with increasing switching rate $\gamma$ the density tends to the no blinking case, i.e, to the situation in which the target is always active; see Eq.~(\ref{eq:mfpt-sr}). For a further discussion of the asymptotic behavior, see Ref.~\onlinecite{mercado2019firsthitting}.
In Fig.~\ref{fig:density}, the Laplace transforms were inverted numerically with the help of Wolfram Mathematica (solid lines), see Eqs.~(\ref{eq:DRqx0A}) and~(\ref{eq:DRqx0I}). 
The curve ``no blinking'' comes from numerically inverting the Laplace transform in the case without blinking~\cite{evans2011diffusion}.
Additional points in Fig.~\ref{fig:density} represent the results of computer simulation; see Sec.~\ref{sec:langevin}.
These numerical results are used here, prior to description of the simulation approach, for completeness of the presentation.

\subsection{Numerics\label{sec:langevin}}
Parallel to the analytical approach presented in Sec.~\ref{sec:analytics}, the process of hitting a blinking target can be studied using stochastic simulations. 
From Eq.~(\ref{eq:langevin}), it is possible to generate an ensemble of trajectories of various lengths from which first passage times can be estimated.
In the next step, from the ensemble of first passage times, it is possible to calculate various system characteristics, e.g., the first passage time distribution, survival probability $\mathcal{S}(t)=1-\mathcal{F}(t)$ (where $\mathcal{F}(t)$ is the cumulative distribution function of first passage times), the mean first passage time, higher moments (if they exist), and other quantifiers that could be difficult to access analytically, e.g., the probability of hitting the target from the left or right.

Equation~(\ref{eq:langevin}) can be approximated by the Euler-Maruyama scheme \cite{higham2001algorithmic,mannella2002}
\begin{equation}
    \label{eq:EMScheme}
x(t+\Delta t) = x(t) +  \sigma \; \sqrt{\Delta t} \; \xi_t,
\end{equation}
where $\xi_t$ represents the sequence of independent, identically distributed random variables following a (standard) normal distribution $N(0,1)$ ($\xi_t \sim \mathcal{N}(0,1)$).
To construct an ensemble of $N$ trajectories $x(t)$ with the initial condition $x(0)=x_0>0$ and the blinking target, Eq.~(\ref{eq:EMScheme}) is integrated until time $\tau$ as defined by Eq.~\eqref{eq:tauR-def}.
In the simulation, we used $N=10^5$ trajectories constructed with the Euler-Maruyama method
(see Eq.~(\ref{eq:EMScheme})),  with the integration time-step 
$\Delta t=10^{-4}$. 
Moreover, the noise strength $\sigma$ is set  $\sigma=\sqrt{2}$, the target is placed at $z=0$, and $x_0\in \{ 0.5, 1 \}$.
In further studies, $z=0$ is considered as the standard target location.

Equation~(\ref{eq:langevin}) along with its discretized version (\ref{eq:EMScheme}) defines the method for trajectory generation.
To fully formulate the model, Eq.~(\ref{eq:EMScheme}) must be extended by the target dynamics and the resetting protocol.
Between resets, position $x(t)$ is approximated by Eq.~(\ref{eq:EMScheme}), while at reset events, the position is set back to $x_0$.
Times between consecutive resets follow the exponential distribution with the rate $r$, i.e., $\phi(t) = r \exp(-rt),$ where $r$ is the (fixed) reset rate.
The resetting protocol does not affect the target dynamics.
The initial state of the target is either deterministic or drawn from a known two point probability distribution $\p(J(0)=A)=1-\p(J(0)=I)$ with the fixed $\p(J(0)=A)$, e.g., $\p(J(0)=A) = \pi_A$.
The target remains in any of the states for an exponentially distributed time determined by switching rates $\alpha$ and $\beta$; see Eq.~(\ref{eq:dn}).
In the $A$ state, the distribution is $\phi(t) = \alpha \exp(-\alpha t)$,  while in the $I$ state  it is $\phi(t) = \beta \exp(-\beta t)$. 
However, during simulations we have considered symmetric case only, i.e., $\alpha=\beta=\gamma$ for which $\pi_A=\pi_I=1/2$.
The core part of the simulation code is summarized in Algorithm~\ref{alg:simulation}, which, for its generality, uses two rates $\alpha$ and $\beta$.
Variables $t$, $t_r$, and $t_g$ are counters used to measure first passage time ($t$), determine resetting events ($t_r$), and change the target state ($t_g$).
In Algorithm~\ref{alg:simulation},  $\mathrm{Exp}(\rho)$ and $\mathcal{N}(0,1)$ represent random numbers following exponential distributions (with the rate $\rho$) and standard normal ($N(0,1)$) densities, while $p(A,I)$ indicates a two point probability distribution from which the initial states of the target are fixed.

The process $x(t)$ is continuous in time; consequently, it may attain values satisfying $x(t) = 0$.
The situation changes once the process is approximated using a finite time step $\Delta t$.
The discretized representation is inherently discrete in time and, owing to the finite precision of floating‑point arithmetic, effectively discrete in space as well.
Under such conditions, a numerical trajectory may not hit the point $x = 0$ exactly but may instead overshoot it from one time step to the next.
To correctly identify such events, we monitor whether the trajectory crosses the point $x = 0$ between two consecutive time steps.
Accordingly, in Algorithm~\ref{alg:simulation}, the condition corresponding to $x = 0$ is operationally defined as
$x(t) \times x(t - \Delta t) \leqslant 0$
which indicates that at times $t$ and $t- \Delta t$ the particle resides on opposite sides of the origin (or at the origin itself).

\begin{algorithm}[H]
\begin{algorithmic}
\For{$i=1 \dots N$}
\State $t \gets 0$
\State $x \gets x_0$
\State
\State $t_r \sim \mathrm{Exp}(r)$
\State
\State $J \sim p(A,I) $
\If{$J=A$} 
\State $t_g \sim \mathrm{Exp}(\alpha)$
\Else
\State $t_g \sim \mathrm{Exp}(\beta)$
\EndIf 
\State
\While{$! (x(t) \neq 0 ~~~\text{and}~~ J(t)=A)$}
\State $x \gets x+ \sigma \sqrt{\Delta t} \;  \mathcal{N}(0,1) $
\State $t \gets t+ \Delta t $
%
\State
\If{$t \geqslant t_r$} 
\State $x \gets x_0 $
\State $t_r \gets t + \mathrm{Exp}(r) $
\EndIf 
%
\State
\If{$t \geqslant t_g$} 
\If{$J=A$} 
\State $J(t) \gets I$
\State $t_g \gets t + \mathrm{Exp}(\beta) $
\Else
\State $J \gets A$
\State $t_g \gets t + \mathrm{Exp}(\alpha) $
\EndIf 
\EndIf 
\State
\EndWhile
\State
\State push\_back($t$)
\State
\EndFor
\end{algorithmic}
\caption{Simulation pseudo-code.}
\label{alg:simulation}
\end{algorithm}

\subsection{Comparison\label{sec:comparison}}

\begin{figure}[h]
\centering
\includegraphics[width=\textwidth]{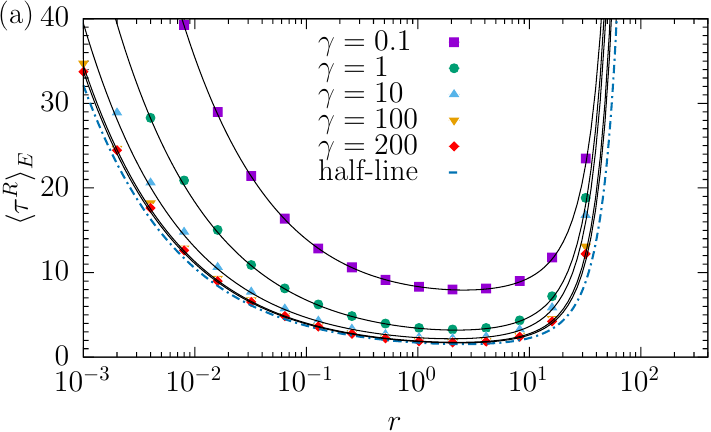}
\includegraphics[width=\textwidth]{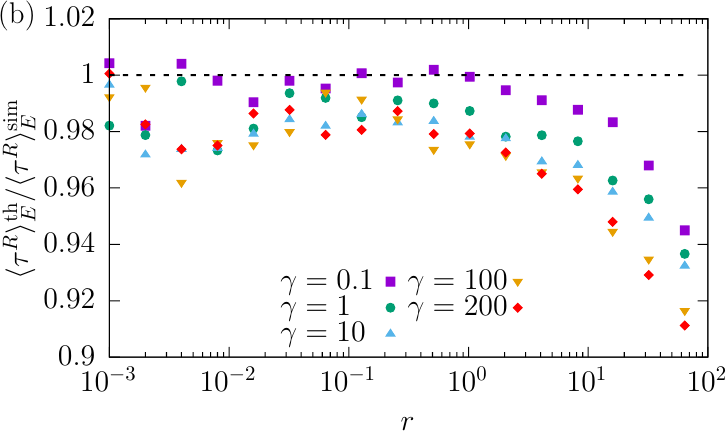}
\caption{The MFPT $\langle \tau^R \rangle_E$ (top panel -- (a)) and ratio of theoretical $\langle \tau^R \rangle_E^{\mathrm{th}}$ and simulation $\langle \tau^R \rangle_E^{\mathrm{sim}}$ results (bottom panel -- (b)).
In the top panel: points represent results of computer simulations, see Sec.~\ref{sec:langevin}, solid lines are analytical formula (Eqs.~(\ref{eq:mean-tauR-A}), (\ref{eq:mean-tauR-I}) in the main text or Eqs.~(\ref{eq:mean-tauR-A-app}) and~(\ref{eq:mean-tauR-I-app}) in Appendix~\ref{sec:bmapp}), and dashed line is the case without blinking (Eq.~(\ref{eq:mfpt-sr})). Model and simulation parameters: $\Delta t = 10^{-4}$, $N=10^5$, $\sigma = \sqrt{2}$, $z=0, x_0 = 1$. 
}
\label{fig:MFPT1}
\end{figure}

\begin{figure}[h]
\centering
\includegraphics[width=\textwidth]{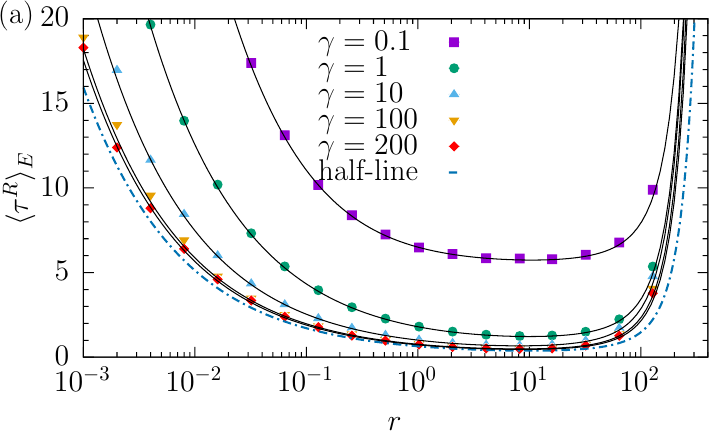}
\includegraphics[width=\textwidth]{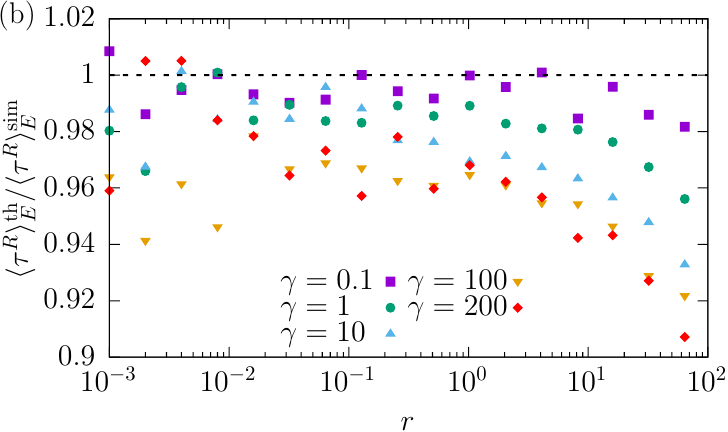}
\caption{The same as in Fig.~\ref{fig:MFPT1} for $z=0, x_0 = 0.5$.}
\label{fig:MFPT05}
\end{figure}

In order to compare the analytical results with numerical simulations, we focused on the exploration of the MFPT $\langle \tau^R \rangle_E$, assuming that the initial state of the target is in equilibrium $E$, see Figs.~\ref{fig:MFPT1} and~\ref{fig:MFPT05}.
Furthermore, it is assumed that the dichotomous process modeling target dynamics is symmetric; therefore, $\alpha=\beta=:\gamma$, see Eq.~(\ref{eq:dn}).
We have performed simulations according to the description provided in Sec.~\ref{sec:langevin} with the initial state of the target drawn from the equilibrium distribution, i.e., $\p(J(0)=A)=\p(J(0)=I)=\frac{1}{2}$.
As discussed above, it is known~\cite{chechkin2018random}, that in the studied BM case (without resetting) $\langle \fpt \rangle = \infty$ and (under resetting) $\langle \fpt^R \rangle < \infty$.
When the target is blinking, we observe an analogous behavior, namely $\langle \tau \rangle = \infty$ and $\langle \tau^R \rangle < \infty$.
The former comes from the fact that for the blinking target $\tau \geqslant \fpt$ and the latter from Eqs.~(\ref{eq:mean-tauR-A})~and~(\ref{eq:mean-tauR-I}).

Fig.~\ref{fig:MFPT1}(a) presents the MFPT for $x(0)=1$ and $z=0$.
The subsequent Fig.~\ref{fig:MFPT05} corresponds to the situation when the distance between the initial condition and the target is reduced to $0.5$.
The points in Figs.~\ref{fig:MFPT1}(a) and~\ref{fig:MFPT05}(a) depict the results of the computer simulation, while the black solid lines represent the analytical results; see Eqs.~(\ref{eq:mean-tauR-A})~and~(\ref{eq:mean-tauR-I}).
The additional dot-dashed line presents the MFPT from the half-line under stochastic resetting; see Eq.~(\ref{eq:mfpt-sr}).
The results obtained for $\langle \tau^R \rangle_E$ from the simulation match the theoretical curves for a wide range of tested reset rates $r$; see Figs.~\ref{fig:MFPT1} and~\ref{fig:MFPT05}.
With the increasing target switching rate $\gamma$, $\langle \tau^R \rangle_E$ approaches the mean first passage time for escape from a half-line under stochastic resetting. 
Such a limiting behavior stems from the fact that for $\gamma \to \infty$ the target is practically always active.
The same behavior was previously observed for this model without stochastic resetting~\cite{mercado2019firsthitting}.

The agreement between computer simulations and theoretical predictions is  further corroborated in the bottom panels of Figs.~\ref{fig:MFPT1} and~\ref{fig:MFPT05}, which depict the ratio between theoretical 
($\langle \tau^R \rangle_E^{\mathrm{th}}$)
and simulation results 
($\langle \tau^R \rangle_E^{\mathrm{sim}}$),
see Appendix~\ref{sec:bmapp} for analytical formulas.
For the selected set of parameters (including the integration time step $\Delta t=10^{-4}$ and the number of repetitions $N=10^5$), errors are typically smaller than 5\%.
Larger errors are recorded for high resetting rates.
The level of agreement can be improved by increasing the number of repetitions; however, for large $r$ (and $\gamma$), it is also necessary to decrease the integration time step $\Delta t$, as the integration time step $\Delta t$ needs to be smaller than other relevant timescales present in the system, i.e., $\Delta t \ll \frac{1}{r}$ and $\Delta t \ll \frac{1}{\gamma}$.

\begin{figure}[h]
    \centering
    \includegraphics[width=0.9\columnwidth]{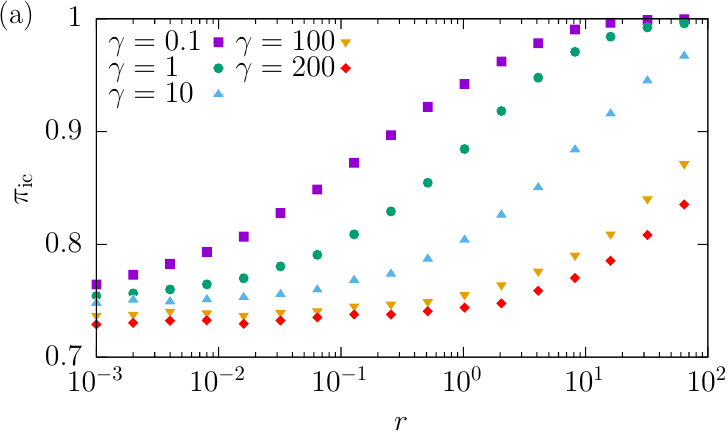}\\
    \includegraphics[width=0.9\columnwidth]{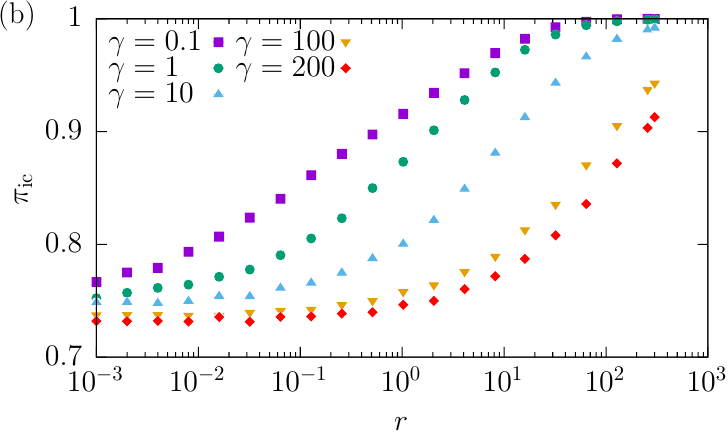}
        \caption{The probability $\pi_{\mathrm{ic}}$ of hitting the target from the initial condition side under stochastic resetting. 
        Various points represents different switching rates $\gamma$.
    In the top panel (a) $x_0=1$, while in the bottom panel (b) $x_0=0.5$.}
    \label{fig:fromright}
\end{figure}

The gate switches between active and inactive states, giving  a particle a chance to pass over the hidden target.
Consequently, the target can be hit from both sides.
The initial condition provides a preference: the dominant fraction of hits is from the side of the initial condition.
Nevertheless, transitions from the opposite  side are also recorded.
Fig.~\ref{fig:fromright} explores the probability $\pi_{\mathrm{ic}}$ of hitting the target from the initial condition side.
With the increasing resetting rate, the motion is restarted more often, resulting in an increase of $\pi_{\mathrm{ic}}$.
Moreover, for fixed resetting rates $r$, the probability $\pi_{\mathrm{ic}}$ increases with the rise in the target switching rate $\gamma$ as it becomes harder to surpass the target.

\begin{figure}[h]
    \centering
    \begin{tabular}{cc}
    \includegraphics[width=0.95\columnwidth]{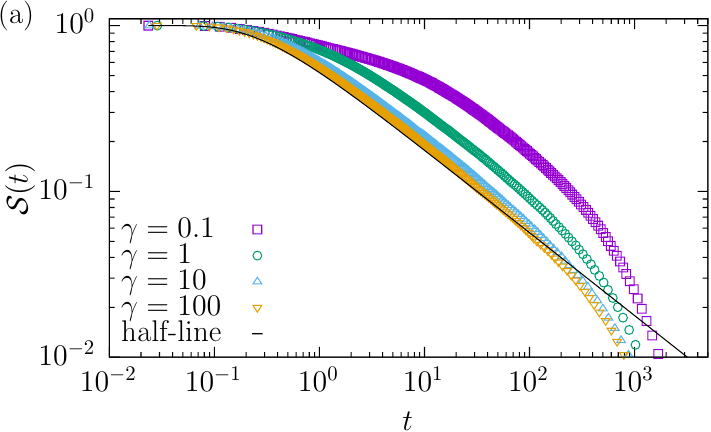}  \\    
    \includegraphics[width=0.95\columnwidth]{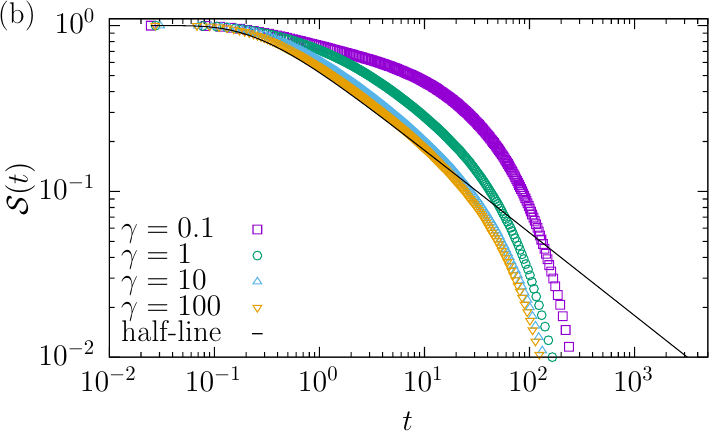} \\
    \includegraphics[width=0.95\columnwidth]{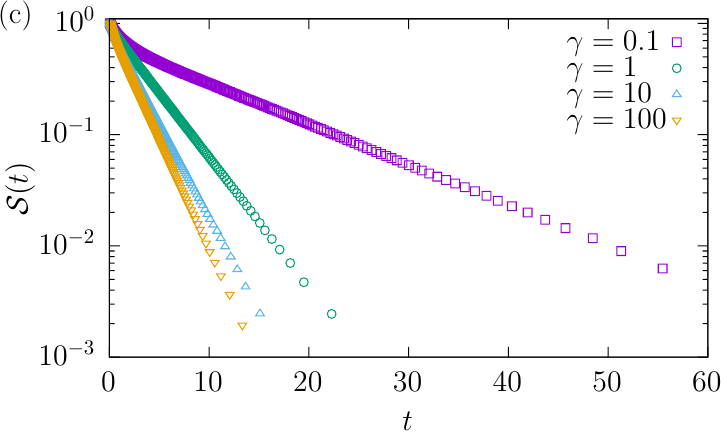}\\
    \end{tabular}
    \caption{Survival probabilities, $\mathcal{S}(t)=1-\mathcal{F}(t)$, with $x_0=1$ and different target switching rates $\gamma$.
    Various panels correspond to various resetting rates:  $r=0.001$ (top panel -- (a)), $r=0.016$ (middle panel -- (b)) and $r=1.024$ (bottom panel -- (c)).
    Solid line in top and middle panel  show survival probability for escape from half-line, see Eq.~(\ref{eq:cdf-hl}).
    Please note that top and middle panels are plot on log-log scale, while the bottom panel on log-lin.}
    \label{fig:surv}
\end{figure}

Finally, we discuss some limiting behaviors of the hitting process. 
In the limit of $r\to 0$ the resetting vanishes. 
Additionally, if one takes $\gamma \to \infty$, the target is practically always in the active state.
Therefore, one can expect that the studied process recovers the escape from the half-line with the always active target, see Eqs.~(\ref{eq:fpt-hl}) and~(\ref{eq:cdf-hl}). 
Indeed, such a limit is evident from numerical simulations.
Fig.~\ref{fig:surv} presents survival probabilities with $x_0=1$ and various values of the resetting rate $r$ and the target switching rate $\gamma$.
Subsequent panels correspond to various resetting rates:  $r=0.001$ (top panel -- (a)), $r=0.016$ (middle panel -- (b)) and $r=1.024$ (bottom panel -- (c)).
The solid line in the top and middle panels shows the survival probability for escape from the half-line; see Eq.~(\ref{eq:cdf-hl}).
Please note that the top and middle panels are plotted in a log-log scale, while the bottom panel is in a log-linear scale.
From Fig.~\ref{fig:surv}, for small fixed $r$, it is visible how $\mathcal{S}(t)$ approaches the power-law, half-line asymptotics; see Eq.~(\ref{eq:cdf-hl}).
Examination of the top panel demonstrates that, with the increasing target switching rate $\gamma$ the survival probability $\mathcal{S}(t)$ nicely approaches the half-line asymptotics.
The position of the point at which $\mathcal{S}(t)$ bends down depends on the resetting rate.
This fact is visible from the comparison of the top and middle panels: $\mathcal{S}(t)$ with $\gamma=100$ for $r=0.001$ follows Eq.~(\ref{eq:cdf-hl}) over a larger range than for $r=0.016$.
Finally, for $r$ large enough, as depicted in the bottom panel, the survival probabilities follow exponential asymptotics.
From Fig.~\ref{fig:surv}, it is clearly visible that the increase in the target switching rate $\gamma$ facilitates escape kinetics.
This is well visible, both at the level of survival probabilities (faster decay) and on the MFPT (smaller values); see Figs.~\ref{fig:MFPT1} and~\ref{fig:MFPT05}.

\section{Summary and conclusions\label{sec:summary}}

The study of escape kinetics in terms of first passage times is a frequent topic in various scientific disciplines. In this work, we investigate the properties of a process that must reach a specific target. However, the target undergoes dichotomous switching between two states and can only be detected when it is in the active state because we assume that the target in the inactive state is invisible.

For both cases, with and without resetting, we provide analytical expressions for the first passage time distributions in the Laplace domain, along with the corresponding mean first passage times. These analytical predictions are compared with numerical simulations, which show excellent agreement. Finally, we discuss the limiting behavior of the model, including the vanishing resetting rate and the high target switching rate.

As an example, we study the 1D model of a free Brownian particle hitting the origin. For this model, the possibility of the target being in the inactive state effectively extends the domain of the process to the entire real line.  Consequently, in the absence of stochastic resetting, the first hitting time has a divergent mean. We show that the introduction of stochastic resetting renders the mean first passage time finite once again.

\section*{Acknowledgments}
The research for this publication has been supported by a grant \emph{Stochastic Navier-Stokes equation in 2D with degenerate noise} from the Strategic Programme Excellence Initiative at Jagiellonian University.
This research received partial support from the funds assigned to AGH by Polish Ministry of Science and Higher Education. 
We gratefully acknowledge Poland’s high-performance computing infrastructure PLGrid (HPC Centers: ACK Cyfronet AGH) for providing computer facilities and support within computational grant no. PLG/2025/018179.
The research is partially supported by the Polish National Science Centre
Grant No. 2023/51/B/ST1/01270.

\appendix
\section{The Derivation of Eq.~(\ref{eq:DqmI})\label{sec:appendix-derivation}}
Here, we present the intermediate steps leading to derivation of Eq.~(\ref{eq:DqmI}), which gives the Laplace transform for a general gated Markov  process without resetting.
As the starting point, we use Eq.~(\ref{eq:Dtx0j0})
\begin{align}
\label{eq:sp1}
    g(t|x_0, j_0) &= f(t| x_0) P^J_{j_0 \to A}(t) + \\ \nonumber
    & + \int_0^t f(t'| x_0) P^J_{j_0 \to I}(t') g(t-t'|z, I) {\rm d}t'~,
\end{align}
and Eq.~(\ref{eq:DtmI})
\begin{align}
\label{eq:sp2}
g(t|z, I) =  \int_0^t \beta e^{-\beta t'} \left[  \int_{-\infty}^{\infty} p(t',x | z) g(t - t'|x ,  A) {\rm d}x \right] {\rm d}t'.
\end{align}
First, we use Eq.~(\ref{eq:sp1}) to replace $g(t - t'|x ,  A)$ in  Eq.~(\ref{eq:sp2})
\begin{align*}
    g(t|z, I) = \\  \nonumber \int_0^t \int_{-\infty}^{\infty}~ \beta e^{-\beta t'}p(t',x | z)\bigg \{ f(t-t'| A) P^J_{A \to A}(t-t') + \\ \nonumber \int_0^{t-t'} f(t''| A) P^J_{A \to I}(t'') g(t-t'-t''|z, I) {\rm d}t''\bigg \} {\rm d}x~{\rm d}t'.
\end{align*}
The only unknown function in the above equation is $g(t|z, I)$, as $g(t|x, A)$ has been eliminated.
In the next step, the Laplace transform is applied to the above equation and the resulting algebraic equation is solved for $\lap g(q|z, I)$.
To do so, we split the RHS into two parts.
The one with the double integral ${\rm d}x~{\rm d}t'$ (call it $I_1(t)$) and the one with the triple integral.
The first part corresponds to Eq.~(\ref{eq:I1q}) of the main text
\[ 
\begin{aligned}
I_1(t) =\\ \int_{-\infty}^{\infty}\int_0^t~ \beta e^{-\beta t'} p(t',x | z) f(t-t'| A) P^J_{A \to A}(t-t'){\rm d}t'~{\rm d}x.
\end{aligned} 
\]
However, to obtain Eq.~(\ref{eq:I1q}), one needs to go to the Laplace space.
In the calculation, we first use the linearity of the Laplace transform to move the space (${\rm d}x$) integral in front, then we recognize a convolution and obtain
\[ \begin{aligned}
\lap I_1(q) =\\ \beta \int_{-\infty}^{\infty} \mathcal{L}\{e^{-\beta t}p(t,x | z)\}(q)~\cdot ~ \mathcal{L}\{f(t| A) P^J_{A \to A}(t)\}(q)~{\rm d}x.
\end{aligned} \]
The final step is to plug in the exact formula for  $P^J_{A \to A}(t)$
\begin{equation}
P^J_{A \to A}(t) = \pi_A + \pi_I e^{-t (\alpha + \beta )}
\end{equation}
and use the shift property $\mathcal{L}\{e^{-\beta t} \phi(t)\}(q) = \mathcal{L}\{\phi(t)\}(q + \beta)$ to get
\begin{eqnarray}
\lap{I_1}(q) & = & \beta \int_{-\infty}^{\infty} \lap p(q + \beta, x | z) \times \\ \nonumber
 && \times \left  [\pi_A \lap f(q|x) + \pi_I \lap f(q + \lambda|x) \right ]{\rm d}x.
\end{eqnarray}
At this stage, our equation has the form
\begin{equation}\label{eq:gwithZ}
\lap g(q|z, I) = \lap I_1(q) + \beta\int_{-\infty}^{\infty} \mathcal{Z}~{\rm d}x.    
\end{equation}
where $\mathcal{Z}$ is given by
\begin{align*}
    \mathcal{Z} = \mathcal{L} \bigg \{ \int_0^t ~ e^{-\beta t'}p(t',x | z) \times \\ \times \Big [ \int_0^{t-t'} f(t''| A) P^J_{A \to I}(t'') g(t-t'-t''|z, I) {\rm d}t''\Big ] {\rm d}t'\bigg \}(q).
\end{align*}
After recognizing two convolutions, we get
\begin{equation}\begin{aligned}
    \mathcal{Z} = \mathcal{L} \Big \{ e^{-\beta t}p(t,x | z) \Big \}(q) \times\\
    \times \mathcal{L} \bigg \{ \int_0^{t} f(t''| A) P^J_{A \to I}(t'') g(t-t''|z, I) {\rm d}t'' \bigg \}(q),
\end{aligned}\end{equation}
and
\begin{equation}\begin{aligned}
    \mathcal{Z} = \mathcal{L} \Big \{ e^{-\beta t}p(t,x | z) \Big \}(q) \times\\
    \times \mathcal{L} \Big \{f(t| A) P^J_{A \to I}(t) \Big \}(q) \times \mathcal{L} \Big \{ g(t|z, I) \Big \}(q).
\end{aligned}\end{equation}
From Eq.~(\ref{eq:gwithZ}) we notice that $\lap g(q|z, I)$ is independent of $x$, thus it can go in front of the space integral
\begin{equation}
\begin{aligned}
\lap g(q|z, I) = \lap I_1(q) +\lap g(q|z, I) \times \\
\times \beta\int_{-\infty}^{\infty} \lap p(q + \beta,x | z) \cdot \mathcal{L} \Big \{f(t| A) P^J_{A \to I}(t) \Big \}(q) ~{\rm d}x.
\end{aligned}
\end{equation}
The above formula justifies the definition of $\lap I_2(q)$ as the part multiplying $\lap g(q|z, I)$ on the RHS.
To get exactly Eq.~(\ref{eq:I2q}), one needs to proceed similarly to $I_1$, i.e., using the formula for $P^J_{A \to I}(t) = \pi_I \left[1-e^{-t (\alpha + \beta )}\right]$ and the Laplace transform shift property.
After those steps, we get
\begin{equation}
    \lap g(q|z, I) = \lap I_1(q) +\lap g(q|z, I) \lap I_2(q),
\end{equation}
which is equivalent to Eq.~(\ref{eq:DqmI}) from main text.
A very similar derivation can be also found in earlier works\cite{szabo1984localized,scher2024continuous}.
We use the formula~(\ref{eq:DqmI}) as a starting point for the discussion of the general gated Markov process under stochastic resetting, see the main text.

\section{Closed formulas for the Brownian Motion \label{sec:bmapp}}

The formulas for $\lap g(q|x_0, j_0)$ ($j_0\in\{A,I\}$) follow from Eqs.~(\ref{eq:DqxA}) and (\ref{eq:DqxI}). 
From these formulas, it is possible to get explicit analytical expressions for BM without resetting
\begin{equation}
\begin{aligned}
    \lap g(q|x_0, A) &= \frac{e^{-|u| \left(\sqrt{2 \gamma +q}+\sqrt{q}\right)}}{2 \left(\gamma +\sqrt{q (2 \gamma +q)}+q\right)} \bigg [ \\
    & \left(\sqrt{q (2 \gamma +q)}+q\right) e^{|u|\sqrt{q}} \\
    & + \left(2 \gamma + \sqrt{q (2 \gamma +q)}+q\right) e^{|u|  \sqrt{2 \gamma +q}}
    \bigg ]
\end{aligned}    
\end{equation}
and
\begin{equation}
\begin{aligned}
    \lap g(q|x_0, I) &= \frac{e^{-|u| \left(\sqrt{2 \gamma +q}+\sqrt{q}\right)}
    }{2 \left(\gamma +\sqrt{q (2 \gamma +q)}+q\right)}
    \bigg [ \\
   & \left(2 \gamma+\sqrt{q (2 \gamma +q)}+q\right) e^{| u|  \sqrt{2 \gamma +q}} \\
   & -\left(\sqrt{q (2 \gamma +q)}+q\right) e^{|u| \sqrt{q}}
   \bigg ]~,
\end{aligned}
\end{equation}
where $u = x_0 \sqrt{2} / \sigma$, $z=0$ and $\alpha=\beta=\gamma$.
The formula for the $\lap g(q|x, j_0)$ can be also found in Ref.~\onlinecite{mercado2019firsthitting}, Eq.~(10) therein, where the authors use a slightly different notation.
For the BM under stochastic resetting, it is possible to write down closed formulas for the Laplace transforms of $\tau^R$ densities and the mean first hitting times $\langle \tau^R \rangle_{A}$ and $\langle \tau^R \rangle_{I}$.
We assume $z=0$, $\alpha=\beta=\gamma$ and, furthermore, define two auxiliary variables $A$ and $B$
\begin{equation}
    A =  \left(\sqrt{(q+r) (2 \gamma +q+r)}+q+r\right)
\end{equation}
and
\begin{equation}
    \begin{aligned}
    B &= q r A e^{|u| \sqrt{q+r}} \\
        &+ 2 q (2 \gamma + q) \left(\gamma + A \right) e^{|u| \left(\sqrt{2\gamma +q+r}+\sqrt{q+r}\right)} \\
        &+ r (2\gamma +q) \left(2 \gamma + A \right) e^{|u| \sqrt{2 \gamma +q+r}}~.
\end{aligned}
\end{equation}
Then, for the BM under resetting, we obtain the Laplace transforms of $\tau^R$ in the following form
\begin{equation}
    \begin{aligned}
        \lap g^R(q|x_0, A) = \frac{1}{B} \bigg [
        &(2\gamma+q) (q+r) \left(2 \gamma +A \right) e^{|u| \sqrt{2 \gamma +q+r}} \\
        &+ q (2\gamma +q+r) A e^{|u| \sqrt{q+r}}
        \bigg ]
    \end{aligned}
\end{equation}
and
\begin{equation}
    \begin{aligned}
        \lap g^R(q|x_0, I) = \frac{1}{B} \bigg [
        &(2 \gamma +q) (q+r) \left(2 \gamma + A \right) e^{|u| \sqrt{2 \gamma +q+r}} \\
        &- q (2 \gamma +q+r) A e^{|u| \sqrt{q+r}}
        \bigg ].
    \end{aligned}
\end{equation}
Additionally we get the MFPT
\begin{align}
\label{eq:mean-tauR-A-app}
    \langle \tau^R \rangle_A &= \frac{e^{| u|  \left(-\sqrt{2 \gamma +r}\right)} }{r \left(2 \gamma
   +\sqrt{r (2 \gamma +r)}+r\right)}
   \Bigg [ \\ \nonumber
   &2 \left(\gamma +\sqrt{r (2 \gamma +r)}+r\right) e^{| u|  \left(\sqrt{2 \gamma +r}+\sqrt{r}\right)} \\ \nonumber
   &- \left(2 \gamma +\sqrt{r (2 \gamma +r)}+r\right) e^{| u|  \sqrt{2 \gamma +r}} \\ \nonumber
   &- \left(\sqrt{r (2 \gamma +r)}+r\right) e^{\sqrt{r} | u| }
   \Bigg ]
\end{align}
and
\begin{align}
\label{eq:mean-tauR-I-app}
    \langle \tau^R \rangle_I &= \frac{e^{| u|  \left(-\sqrt{2 \gamma +r}\right)}}{\gamma r \left(2 \gamma +\sqrt{r (2 \gamma +r)}+r\right)}
   \Bigg [\\ \nonumber
   &\left(\sqrt{r (2 \gamma+r)}+r\right) (\gamma +r) e^{\sqrt{r} | u| } \\ \nonumber
   &+ 2 \gamma \left(\gamma +\sqrt{r (2 \gamma+r)}+r\right) e^{| u| \left(\sqrt{2 \gamma +r}+\sqrt{r}\right)} \\ \nonumber
   &- \gamma \left(2 \gamma +\sqrt{r (2 \gamma +r)}+r\right) e^{| u|  \sqrt{2 \gamma +r}}
   \Bigg ].
\end{align}
Eqs.~(\ref{eq:mean-tauR-A-app}) and (\ref{eq:mean-tauR-I-app}) are obtained from Eqs.~(\ref{eq:mean-tauR-A}) and~(\ref{eq:mean-tauR-I}) after inserting $\lap g(q|x, j_0)$, therefore the intermediate step of calculating $\lap g^R(q|x_0, j_0)$ is not necessary to get MFPT.
Eqs.~(\ref{eq:mean-tauR-A-app}) and (\ref{eq:mean-tauR-I-app}) agree with the results of Ref.~\onlinecite{mercado2021search}, which studies BM under stochastic resetting.
In comparison to earlier studies\cite{mercado2019firsthitting,mercado2021search,biswas2023rate} the approach used within the current manuscript is more general and can be applied not only for BM.
It accommodates a broad class of stochastic processes, including both standard BM\cite{mercado2019firsthitting,mercado2021search} and BM with a drift\cite{biswas2023rate} among others.

\section*{References}
\bibliography{core-bibliography.bib}

\end{document}